\documentclass[amsmath,amssymb,aps,twocolumn,prb,longbibliography,floatfix,superscriptaddress]{revtex4-2}
\usepackage{graphicx,bm,times}
\usepackage[utf8]{inputenc}
\usepackage{braket}
\usepackage{amsfonts}
\usepackage{color}

\usepackage[breaklinks]{hyperref}
\hypersetup{
    pdfstartview={FitH},
    colorlinks=true,
    linkcolor=blue,
    citecolor=blue,
    urlcolor=blue
}

\begin{document}

\title{Impact of electron correlations on the $\mathbf{k}$-resolved electronic structure of PdCrO$_{2}$ revealed by Compton scattering}

\author{A.~D.~N.~James}
\affiliation{H.~H.~Wills Physics Laboratory,
University of Bristol, Tyndall Avenue, Bristol, BS8 1TL, United Kingdom}
\author{D.~Billington}
\affiliation{School of Physics and Astronomy, Cardiff University,
Queen’s Building, The Parade, Cardiff, CF24 3AA, United Kingdom}
\author{S.~B.~Dugdale}
\affiliation{H.~H.~Wills Physics Laboratory,
University of Bristol, Tyndall Avenue, Bristol, BS8 1TL, United Kingdom}

\date{\today}

\begin{abstract}
Delafossite PdCrO$_2$ is an intriguing material which displays nearly-free electron and Mott insulating behaviour in different layers. Both angle-resolved photoemission spectroscopy (ARPES) and Compton scattering measurements have established a hexagonal Fermi surface in the material's paramagnetic phase. However, the Compton experiment detected an additional structure in the projected occupancy which was originally interpreted as an additional Fermi surface feature not seen by ARPES. Here, we revisit this interpretation of the Compton data. State-of-the-art density functional theory (DFT) with dynamical mean field theory (DMFT), the so-called DFT+DMFT method, predicts the Mott insulating state along with a single hexagonal Fermi surface in excellent agreement with ARPES and Compton. However, DFT+DMFT fails to predict the intensity of the additional spectral weight feature observed in the Compton data. We infer that this discrepancy may arise from the DFT+DMFT not being able to correctly predict certain features in the shape and dispersion of the unoccupied quasiparticle band near the Fermi level. Therefore, a theoretical description beyond our DFT+DMFT model is needed to incorporate vital electron interactions, such as  inter-layer electron coupling interactions which for PdCrO$_2$ gives rise to the Kondo-like so-called intertwined excitation. 

\end{abstract}

\maketitle

\section{Introduction}

Interest has grown over the last few decades in layered triangular-lattice delafossite materials with chemical formula $AB$O$_2$ ($A$ = Pt, Pd, Ag or Cu, and $B$ = Cr, Co, Fe, Rh, Al, Ga, Sc, In or Tl). This interest in metallic delafossites was sparked by reports from Tanaka ~\textit{et al.}~\cite{Tanaka.1996,Tanaka.1997} of strongly anisotropic conductivity in their PdCoO$_2$ and PtCoO$_2$ single crystals. Previous measurements of PtCoO$_2$ displayed an extremely low in-plane room temperature resistivity of 3~$\mu\Omega$cm, a value comparable to elemental Cu \cite{shannon1971chemistry,Mackenzie_2017}. This led to the emergence of a new field of research into these materials~\cite{Mackenzie_2017}. The PdCrO$_2$ compound also has the anticipated anisotropic conductivity~\cite{Takatsu.2010.2}, but displays an antiferromagnetic phase below its N\'eel temperature, $T_{\rm N}=37.5$~K, above which the local Cr$^{3+}$ ($S=3/2$) electron spins in the CrO$_2$ layers are frustrated. Within the antiferromagnetic phase this frustration is relieved, resulting in the local spins ordering with a rotation of 120$^{\circ}$ between adjacent sites \cite{Takatsu.2009,Takatsu.2010.2,Takatsu.2014,billington2015magnetic}. The observation of this ordered state offers an opportunity to study the coupling between nearly-free electrons and (frustrated) local electron spins in a frustrated antiferromagnet. This interest in PdCrO$_2$ has led to further experimental characterisation of this material, leading to the discovery of an unconventional anomalous Hall effect~\cite{Takatsu.2010,Takatsu.2014}, and a reconstructed Fermi surface within the (smaller) antiferromagnetic Brillouin zone, measured by both angle resolved photoemission spectroscopy (ARPES)~\cite{Sobota.2013,Noh.2014,Sunko.2020} and quantum oscillations~\cite{Ok.2013,PhysRevB.92.014425}. For PdCrO$_2$, it has been implicitly assumed that electron correlations are the driving force for the antiferromagnetic state, and hence why the behaviour of the CrO$_2$ layer has been described with the concept of local moments~\cite{Mackenzie_2017}. 

Within the paramagnetic phase, both ARPES~\cite{Sobota.2013,Noh.2014} and Compton scattering~\cite{billington2015magnetic} measurements were performed to determine the Fermi surface geometry. ARPES measures the energies of the emitted photoelectrons from the sample surface together with their angle of emission such that the quasiparticle energy and its dispersion with (crystal) momentum up to and including the Fermi energy can be extracted. The ARPES spectra show both the ground and excited states of the electronic structure and measurements are sensitive to the surface and matrix element effects~\cite{Bansil_2004}. Compton scattering experiments probe the bulk ground-state electronic structure through its electron momentum distribution~\cite{barbiellini:13} by measuring so-called Compton profiles which are the doubly projected electron momentum densities (EMDs)~\cite{Dugdale_2014}. The EMD is the electron density distribution in real momentum, $\mathbf{p}$, which can be directly related to the electron occupancy by folding the EMD back into the first Brillouin zone (the Lock-Crisp-West (LCW) theorem~\cite{Lock_1973}) to recover the full translational symmetry of the reciprocal lattice. This folded EMD is now a function of the crystal momentum, $\mathbf{k}$. Electron occupancy in $\mathbf{k}$-space is influenced by temperature, site disorder, and many-body electron correlations. The step changes in the occupancy can be used to determine the Fermi wave-vectors (and hence Fermi surface) even in materials which are either inaccessible by or challenging for other techniques. Such materials include highly chemically-disordered alloys~\cite{PhysRevLett.124.046402} which have short electronic mean-free paths. Evidently, ARPES and Compton scattering probe different aspects of the electronic structure. The Fermi surface may be extracted from either the  $\mathbf{k}$-resolved photoelectron dispersion around the Fermi energy measured by ARPES, or the changes in the occupation derived from the Compton data. 

Both ARPES and Compton scattering confirmed the presence of the hexagonal Fermi surface, but the Compton experiments clearly showed an additional contribution to the projected electron occupancy around the corners of the (projected) Brillouin zone. In an effort to understand their result, Billington \textit{et al.}~\cite{billington2015magnetic} performed density functional theory (DFT) calculations from which they concluded that at least two Fermi surface sheets were required to describe all the features in the $\mathbf{k}$-space occupancy. This led to speculation by Billington \textit{et al.} that what appeared to be an additional Fermi surface sheet observed in the bulk-sensitive Compton experiments but not in the ARPES might be due to some combination of the surface not being representative of the bulk or unfavourable matrix elements. Although Ong \textit{et al.}~\cite{PhysRevB.85.134403} had shown that the DFT magnetic structure of PdCrO$_2$ was three dimensional, at the time of the study by Billington \textit{et al.} there were no published DFT calculations of non-magnetic PdCrO$_2$ in opposition to their two Fermi surface model. 

This interpretation of the Compton data was subsequently critically examined by Mackenzie~\cite{Mackenzie_2017} who argued that it did not take into account the existence of the Mott insulating state in the CrO$_2$ layers (the existence of which is supported by several experiments). However, these arguments do not explain the extra features in the occupation number measured by the Compton scattering. Recent calculations combining DFT with dynamical mean field theory (DFT+DMFT)~\cite{Lech.2018,Sunko.2020,Lechermann.2021} showed that the Mott insulating state in the CrO$_2$ layers is a natural consequence of the inclusion of the local dynamical electron correlations. Also, DFT+DMFT naturally includes paramagnetic electron correlations within the DMFT part~\cite{ABRIKOSOV201685} which is vital for this frustrated antiferromagnetic material. Therefore, the interpretation of the results from the Compton experiment warrants further investigation in order to reconcile it with the picture of local moments within the Mott insulating CrO$_2$ layers and to help resolve the inconsistent conclusions about the Fermiology from the different measurements. 

In light of the recent PdCrO$_2$ DFT+DMFT calculations, it is necessary to first reproduce them in order to then determine the DFT+DMFT EMD using the recent technique implemented by James \textit{et al.}~\cite{james2020magnetic}. From such calculations, a comparison with the Compton scattering experiment, primarily the projected occupation, could be made. With respect to a non-interacting prediction, the inclusion of many-body correlations (such as that predicted by Fermi liquid theory~\cite{giuliani_vignale_2005}) generally leads to a redistribution and apparent smearing in $\mathbf{k}$-space of the occupation around the Fermi wave-vector. The presence of the Mott insulating CrO$_2$ layers in the previous DFT+DMFT predictions lead to significant changes to the shape and dispersion of the quasiparticle bands which also means that there would be significant changes to the occupation, which the Compton scattering will be sensitive to. Hence it is important to use DFT+DMFT to determine whether the predicted electronic structure with these Mott insulating CrO$_{2}$ layers are compatible with the electron occupancy as measured by Compton scattering.

In this study, we revisit the interpretation of the Compton scattering experimental results and compare them with the corresponding quantities calculated from the non-magnetic DFT and paramagnetic DFT+DMFT methods. Here, we see that the size and shape of the predicted DFT+DMFT hexagonal Fermi surface is in excellent agreement with the ARPES~\cite{Sobota.2013,Noh.2014,Sunko.2020}, quantum oscillations~\cite{Ok.2013,PhysRevB.92.014425}, and Compton measurements~\cite{billington2015magnetic}. However, there are still discrepancies between the experimental Compton data and the DFT+DMFT calculations around the corners of the Brillouin zone. These discrepancies can be reduced (but not eliminated) in the DFT+DMFT calculation by artificially (and unphysical) broadening the unoccupied quasiparticle band just above the Fermi level around the corners of the Brillouin zone. This suggests that changes to both the shape and dispersion of that quasiparticle band are required, most likely driven by certain electron correlation effects which theories beyond our DFT+DMFT would possibly capture, such as inter-layer electron coupling interactions which gives rise to the previously observed (Kondo-like) so-called intertwined excitation in Ref.~\cite{Sunko.2020} which is a convolution of the charge spectrum of the metallic layer and the spin susceptibility of the Mott insulating layer.

\begin{figure*}[t!]
 \centerline{\includegraphics[width=0.995\linewidth]{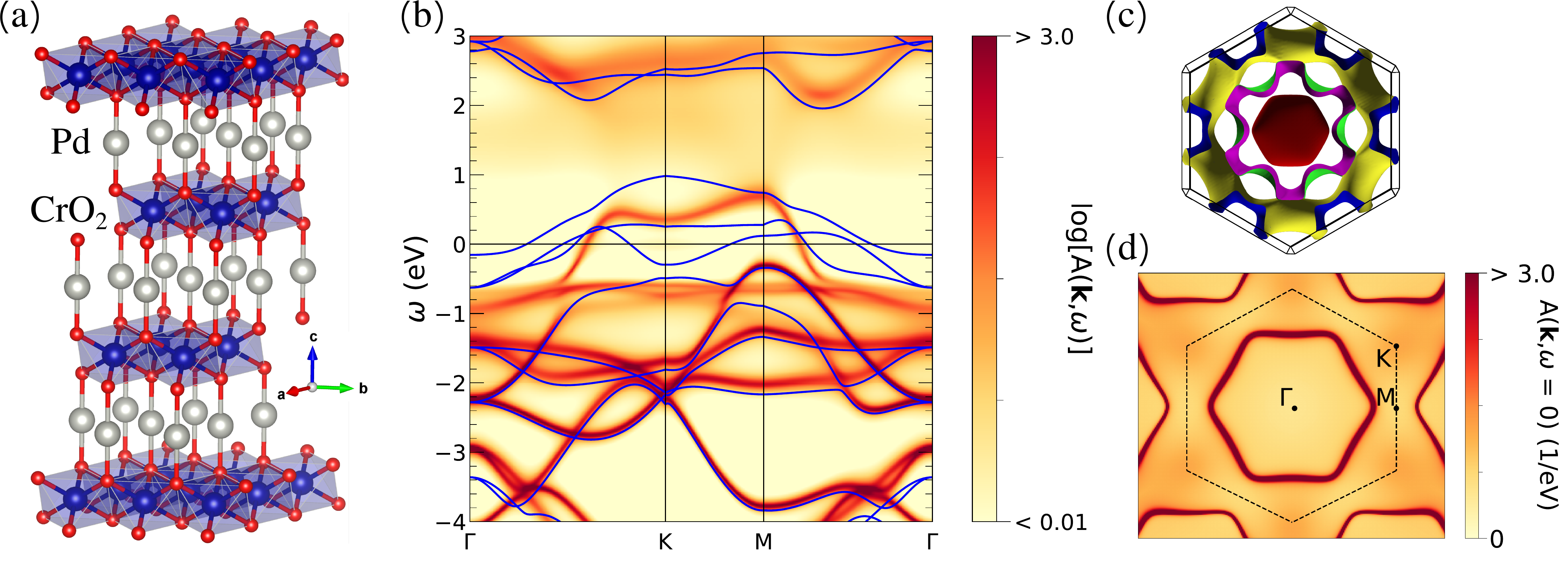}} 
 \caption{(a) The PdCrO$_2$ delafossite structure showing the triangular-lattice Pd and CrO$_2$ layers. (b) The logarithm of the DFT+DMFT spectral spectral function $A(\textbf{k}$, $\omega)$ overlaid with the DFT band structure (blue solid lines). These have been evaluated along a path connecting points within the $k_z = 0$ plane of the Brillouin zone of PdCrO$_2$. The points $\Gamma$ and $K$ are high-symmetry points, with $K$ on the Brillouin zone boundary of the primitive rhombohedral cell. While $M$ is not a high-symmetry point in that Brillouin zone, it is used here with reference to the equivalent point in a simple hexagonal Brillouin zone, as in previous work \protect\cite{Lech.2018,Lechermann.2021,Sunko.2020}. Here, the changes to the Fermi surface between these theoretical methods are most prominent. (c) The three DFT Fermi surface sheets in the rhombohedral Brillouin zone and (d) the DFT+DMFT hexagonal Fermi surface (given by the spectral function evaluated at the Fermi level where $\omega=0$ eV) in the same $k_z$ plane as described in (b). Note that there is distinguishable spectral weight at $K$ with respect to the $\Gamma$ and $M$ points.} 
\label{dmftband}
\end{figure*}

\section{Methods}
We have used the full potential augmented plane-wave plus local orbitals (APW+lo) {\sc elk} code~\cite{elk} in combination with the toolbox for research on interacting quantum systems (TRIQS) library~\cite{triqs}. This so-called {\sc elk}+TRIQS DFT+DMFT framework is described in Ref.~\cite{james_2021}. Further discussion of interfacing the APW+lo DFT basis with the DMFT Anderson's impurity model is found in Ref.~\cite{aichhorn2009}. The PdCrO$_{2}$ delafossite structure is shown in Fig.~\ref{dmftband}~(a) and the lattice parameters of the conventional (hexagonal) unit cell are $a$ = 2.929 \AA, $c$ = 18.093 \AA~\cite{Takatsu.2014} with the Pd--O distance, $d_{{\rm Pd}-{\rm O}} = 0.11c$. The DFT calculation used the Perdew-Burke-Ernzerhof (PBE) generalized gradient approximation (GGA) for the exchange-correlation functional~\cite{PhysRevLett.77.3865} and was converged on a $32 \times 32 \times 16 $ Monkhorst-Pack $\mathbf{k}$-mesh of 2601 irreducible $\mathbf{k}$-points in the irreducible Brillouin zone. We used the all-electron full-potential APW+lo DFT method instead of the pseudo-potential plane-wave approach used in Refs.~\cite{Lech.2018,Lechermann.2021}. The DFT outputs were interfaced to the TRIQS/DFTTools application of the TRIQS library~\cite{aichhorn2016} by constructing Wannier projectors, as described in Ref.~\cite{james_2021}, for all the Cr $3d$-states within a correlated energy window of $[-8.5, 3]$~eV relative to the Fermi level. 

The paramagnetic DMFT part of the DFT+DMFT calculation was implemented using the continuous-time quantum Monte Carlo (CT-QMC) solver within the TRIQS/CTHYB application~\cite{seth2016} with the Slater interaction Hamiltonian parameterised by the Hubbard interaction $U = 3.0$~eV and Hund exchange interaction $J = 0.7$~eV, unless otherwise specified. These $U$ and $J$ values are similar to those used in previous calculations of PdCrO$_2$ ~\cite{Lech.2018,Sunko.2020,Lechermann.2021}, and other CrO$_2$ compounds~\cite{Lechermann.2021,PhysRevLett.80.4305}. We approximated the double counting in the fully localised limit in line with the previous DFT+DMFT calculations~\cite{Lech.2018,Sunko.2020,Lechermann.2021}. Our DFT+DMFT approach slightly differs from previous DFT+DMFT calculations where different correlated energy windows were used and either the Hubbard-Kanamori interaction Hamiltonian~\cite{Lech.2018,Lechermann.2021} or the Hubbard I approximation for the impurity solver~\cite{Sunko.2020} were chosen. We note that our DFT+DMFT calculations are paramagnetic with no overall ordered moment. The Cr $3d$ orbitals were diagonalised from the complex spherical harmonic basis into the diagonal trigonal basis (obtained by diagonalising the orbital density matrix) resulting in the three sets of non-degenerate orbitals, namely the two doubly degenerate $e^{\prime}_g$ and $e_g$ orbitals along with the single $a_{1g}$ orbital, in agreement with Ref.~\cite{Lech.2018}. We used the fully-charge-self-consistent (FCSC) DFT+DMFT method with a total of $8.4\times10^{7}$ Monte Carlo sweeps within the impurity solver for each DMFT cycle. An inverse temperature $\beta = 40$ eV$^{-1}$ ($\sim$~290~K) was used which is similar to the (room) temperature of the Compton scattering experiments. The spectral functions were calculated by analytically continuing the DMFT self-energy obtained from the {\em LineFitAnalyzer} technique of the maximum entropy analytic continuation method implemented within the TRIQS/Maxent application \cite{PhysRevB.96.155128}. 

For the DFT and DFT+DMFT EMD calculations, we used the method of Ernsting \textit{et al.}~\cite{ernsting2014calculating} together with the DFT+DMFT L\"owdin-type basis electron wave functions and occupation numbers determined by the method described in Ref.~\cite{james2020magnetic}. A maximum momentum cut-off of 16.0 a.u. was used. We emphasise that the EMD related results do not use analytic continuation so they do not suffer from its associated complications. We concentrate on the projected EMD for comparisons with the experimental Compton data. To compare with the experimental 2D occupancy in Ref.~\cite{billington2015magnetic}, which directly relates to the electron occupation, the calculated EMDs were first projected along the $k_z$-axis (parallel to the $c$-axis of the conventional unit cell) and this projected 2D EMD was then convoluted with a 2D Gaussian function with a full-width-at-half-maximum of 0.106 a.u. approximating the effect of the finite Compton scattering experimental momentum resolution~\cite{billington2015magnetic}. These convoluted EMDs are subsequently folded back into the first Brillouin zone, via the LCW theorem, producing the theoretical 2D projected occupancy. The Compton profiles, $J(p_{z})$, which are double-projections of the EMD, were evaluated along the experimental scattering vectors (which for convenience are conventionally referred to as being along $p_z$ in the local coordinate system),
\begin{equation}
J(p_{z})=\iint \rho(\mathbf{p}) \mathrm{d} p_{x} \mathrm{d} p_{y}, \label{eq:J}
\end{equation}
where $\rho(\mathbf{p})$ is the 3D EMD. The so-called directional differences, which are the differences between Compton profiles resolved along different crystallographic directions, were calculated so that they could be compared to the experimental ones.

\section{Results}

\begin{figure}[t!]
 \centerline{\includegraphics[width=\linewidth]{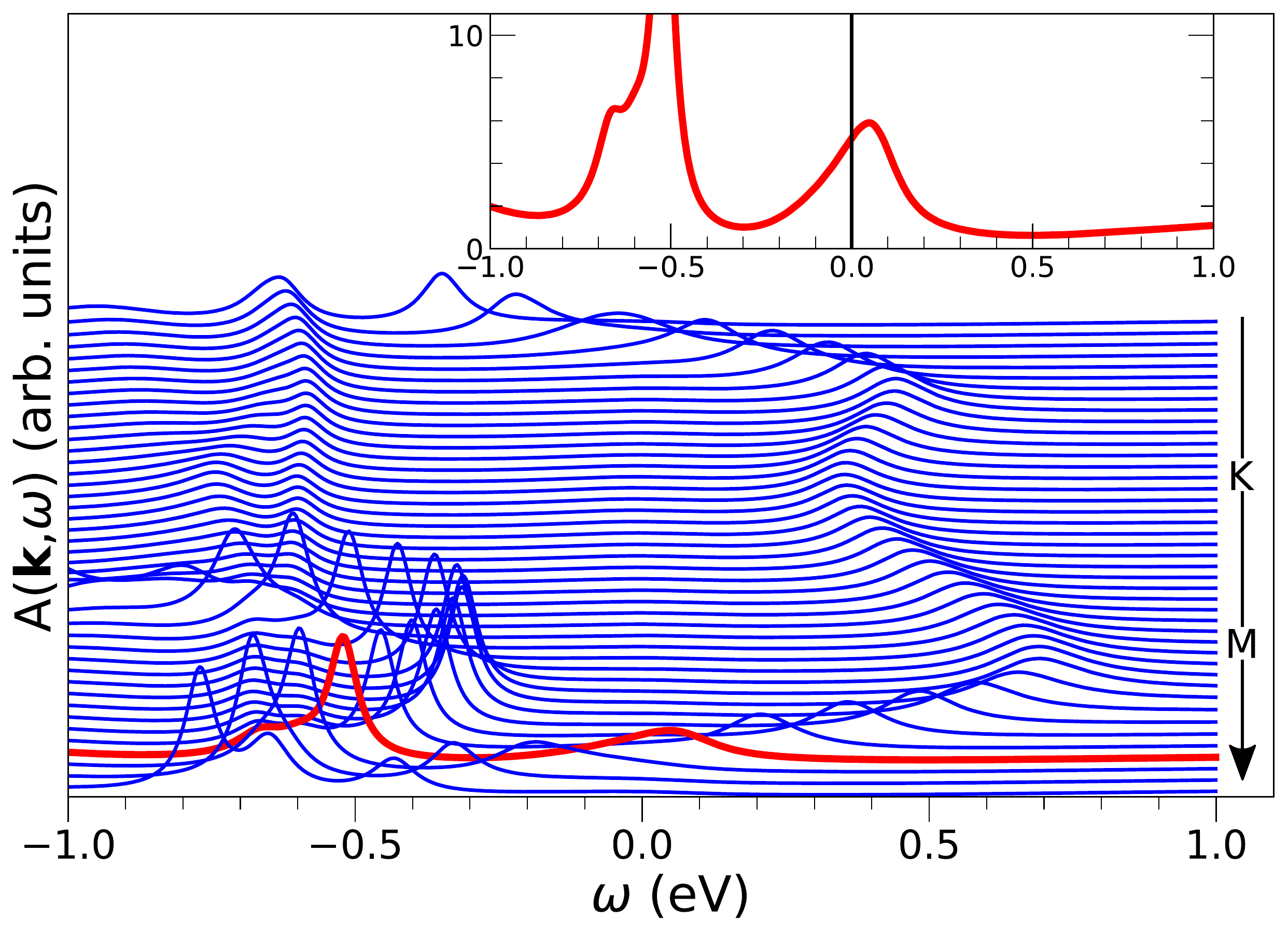}}
 \caption{The DFT+DMFT (fixed $U=3.0$~eV and $J=0.7$~eV) spectral function $A(\textbf{k}$, $\omega)$ plotted in the style of ARPES energy distribution curves (EDCs). This shows the spectral function dispersion around the Fermi level along a portion of the path of Fig.~\ref{dmftband}~(b) focusing on the Pd quasiparticle conduction band crossing the Fermi level ($\omega=0$ eV). The inset reveals structure in the spectral function evaluated at a $\mathbf{k}$-point between $M$ to $\Gamma$ which is highlighted in red in the EDCs. The axes of the inset are the same as the main figure. The Pd quasiparticle conduction band centre is just above the Fermi level, but there is spectral weight from this Lorentzian-like quasiparticle band spectral function crossing the Fermi level and this occupied spectral weight will contribute to the occupation distribution. This occupied weight is referred to as spectral weight spillage across the Fermi level.} 
\label{EDC}
\end{figure}

\begin{figure*}[t!]
 \centerline{\includegraphics[width=0.85\linewidth]{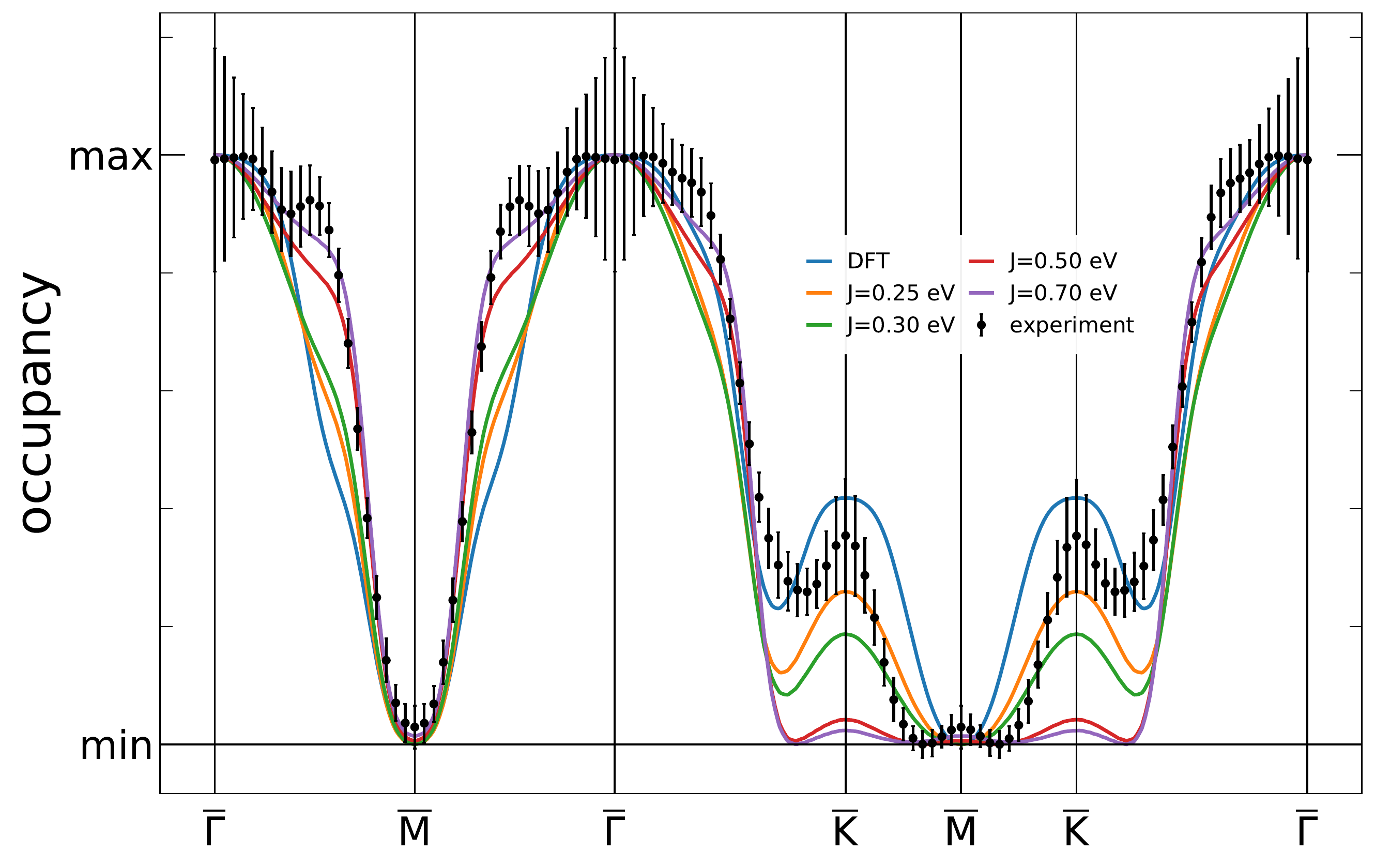}}
 \caption{The 2D occupancy (projected along the $k_z$-axis) plotted along the projected bulk high-symmetry directions (denoted with overlines) for DFT and DFT+DMFT with different values of the Hund exchange $J$ at fixed Hubbard $U = 3.0$~eV. The theoretical projected EMDs were convoluted with a two dimensional Gaussian (full-width-at-half-maximum $=0.106$ a.u.) to approximate the effect of the finite experimental momentum resolution prior to calculating the occupancy. The experimental data are from Ref.~\cite{billington2015magnetic}. Varying $J$ explores the changes to the electronic structure passing through the Mott transition of the Cr states, with the Mott insulating state occurring for $J>0.25$~eV.} 
\label{trans_occ}
\end{figure*}

Fig.~\ref{dmftband}~(b) shows the DFT band structure and DFT+DMFT spectral function plotted along the high symmetry directions in the $k_z = 0$ plane. The DFT and DFT+DMFT results show good agreement with previous studies~\cite{Lech.2018,Sunko.2020,Lechermann.2021}. We note that our spin-orbit coupling DFT calculation differs to that presented in Ref.~\cite{billington2015magnetic}, even though those previously published results are reproducible with the same version (2.2.9) of {\sc elk}. The lack of reproducibility of the Ref.~\cite{billington2015magnetic} ground state with the current version of {\sc elk} suggests that there was some problem with that calculation in version 2.2.9 (which has been fixed in later versions) which coincidentally gave convincing agreement between the reported electronic structure and experimental Compton data. In agreement with the other previously reported DFT and DFT+DMFT predictions, the hybridised Pd $4d$ and Cr $3d$ DFT bands which lie around the Fermi level and which contribute to the DFT Fermi surface shown in Fig.~\ref{dmftband}~(c) drastically redistribute, with the Cr $3d$ dominant bands now insulating in DFT+DMFT due to the formation of a Mott insulating state within the CrO$_2$ layers which arises from the strong local electron correlations on the Cr site. The remaining quasiparticle band which crosses the Fermi level in DFT+DMFT $A(\textbf{k}$, $\omega)$ is now predominantly Pd $4d$ in character and forms the hexagonal Fermi surface sheet shown in Fig.~\ref{dmftband}~(d), in excellent agreement with that observed in the paramagnetic ARPES \cite{Noh.2014} measurements. There are also incoherent, non-dispersive Hubbard-like bands, shown in Fig.~\ref{dmftband}~(b) centred around $\pm 1.5$ eV, which arise from the Mott insulating Cr states. We note that the DFT+DMFT spectral function in Fig.~\ref{dmftband}~(d) shows significant spectral weight around the $K$ point which is also seen in previous DFT+DMFT calculations by Lechermann~\cite{Lech.2018}.

To help illustrate certain concepts which link the spectral function to the occupation distribution (required for subsequent discussions), we have included the DFT+DMFT spectral function around the Fermi level in Fig.~\ref{EDC}, plotted in the style of ARPES energy distribution curves (EDCs). The spectral function of the Pd dominant quasiparticle conduction band is seen to be broader and have a smaller amount of spectral weight than the inverted parabolic quasiparticle band around $M$ which peaks at about $-0.5$ eV (which is also shown in the inset of Fig.~\ref{EDC}). The inset shows that at a particular $\mathbf{k}$-point between $M$ and $\Gamma$ the Pd quasiparticle conduction band centre is just above the Fermi level which of course means that there is no Fermi surface at this wave-vector. However, owing to the finite width of the spectral function around the quasiparticle peaks (which arises from the finite lifetime linked to the imaginary part of the DMFT self-energy), there is a portion of the spectral function tail which crosses the Fermi level and is consequently occupied. This occupied portion of the Pd quasiparticle conduction band contributes to the EMD and will be seen in the electron occupancy measured by Compton scattering. Conversely, if the band centre (quasiparticle peak) were below the Fermi level, but the higher energy tail crosses the Fermi level, then that quasiparticle band will have a reduced contribution to the occupation at that ${\bf k}$-point with respect to a fully occupied quasiparticle band. We refer to the spectral weight from the quasiparticle band tails crossing the Fermi level as spectral weight spillage. The spectral weight spillage will be dependent on factors which influence the finite width (inverse lifetime) of the (quasiparticle) peaks in the spectral function. In the DFT picture within the Green's function formalism, the typical DFT spectral function would be a series of Lorentzian-like functions corresponding to the DFT bands and most likely have small widths relating to the temperature used in the calculation. The corresponding occupation distribution will therefore have contributions from the fully occupied spectral function below the Fermi level and from spectral weight spillage. The common consequence of spectral weight spillage contribution in DFT is the apparent smearing of the occupation distribution in (crystal) momentum around the Fermi wave-vector (which is temperature dependent because of the temperature dependence of the spectral weight spillage). The effects of spectral weight spillages on the occupation distribution are often less prominent in DFT but have been seen for DFT bands grazing the Fermi level such as in ZrZn$_2$~\cite{Chen.2022} and in highly compositionally disordered systems~\cite{PhysRevLett.124.046402}.

\begin{figure}[t!]
 \centerline{\includegraphics[width=\linewidth]{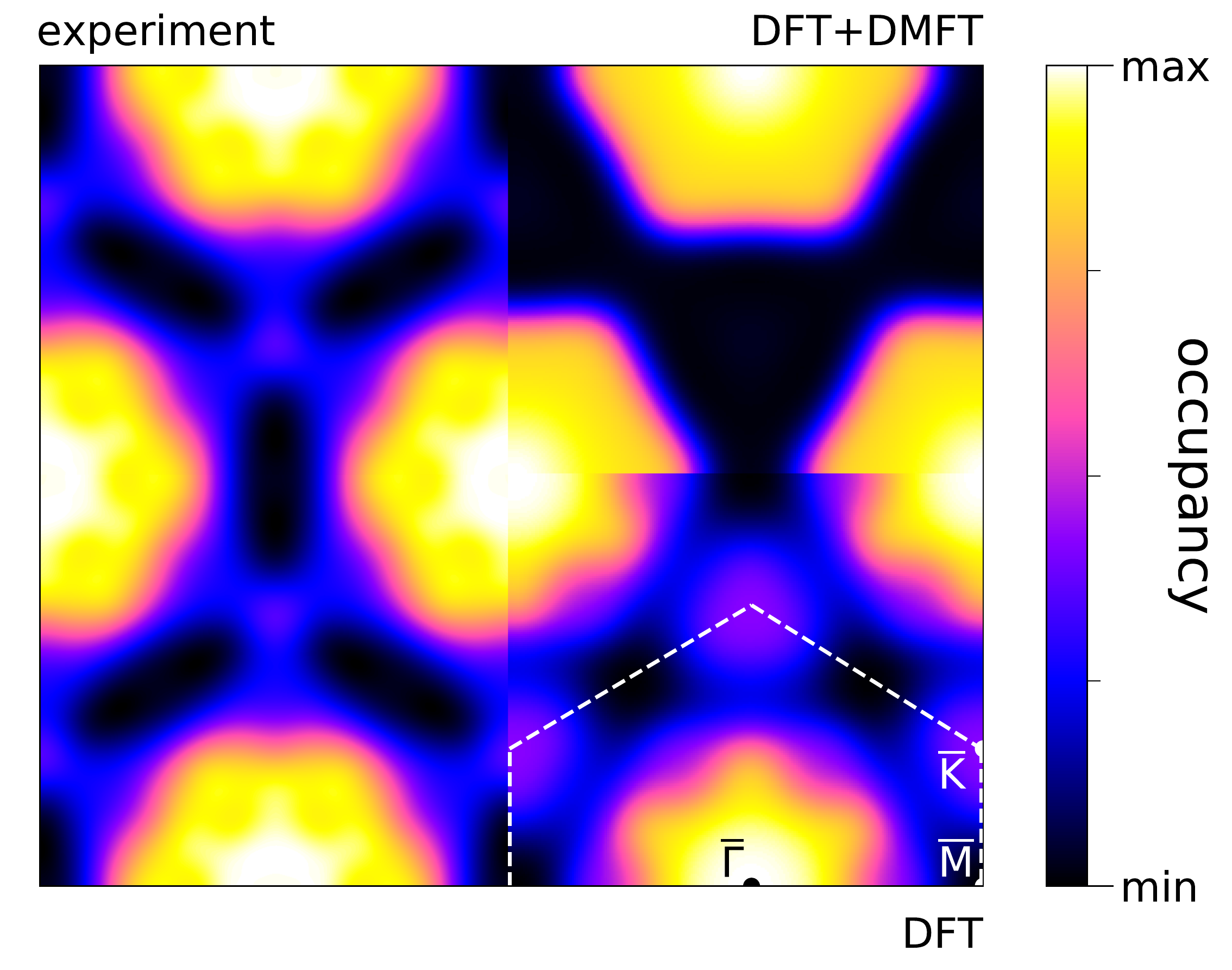}}
 \caption{The 2D (projected along the $k_z$-axis) occupancy in the 2D hexagonal Brillouin zone. The left hand side shows the experimental data, whereas each quadrant on the right hand side represents a different calculation, as indicated. The theoretical two dimensional EMDs were convoluted as described in the Fig.~\ref{trans_occ} caption. The DFT quandrant includes the Brillouin zone boundary as well as the projected 2D high symmetry points (denoted with overlines). The experimental data are from Ref.~\cite{billington2015magnetic}.} 
\label{2docc}
\end{figure}

The 2D projected occupancy (along the projected bulk high symmetry path used in Ref.~\cite{billington2015magnetic}) determined from the DFT and DFT+DMFT calculated EMDs, together with the the experimental 2D occupancy, are shown in Fig.~\ref{trans_occ}. Here, we see that the agreement in the DFT+DMFT ($U=3.0$~eV, $J=0.7$ eV) 2D projected occupancy significantly improves along the $\overline{\Gamma}$ to $\overline{M}$ direction compared to the DFT results, with there being a single step along this direction in the DFT+DMFT compared to the smoothed shoulder predicted by the DFT. The location of this single step along $\overline{\Gamma}$ to $\overline{M}$ gives the Fermi wave-vector of the hexagonal Fermi surface sheet along this direction. We can also extract the Fermi wave-vector of the hexagonal Fermi surface sheet along the $\overline{\Gamma}$ to $\overline{K}$ from the location of the largest change in the projected occupation. The DFT+DMFT projected occupation which relates to hexagonal Fermi surface sheet along with the region it encompasses (see Fig.~\ref{2docc}) is in excellent agreement with the Compton data. We find the occupied fraction of the Brillouin zone associated with DFT+DMFT hexagonal Fermi surface is approximately equal to one half, which is in excellent agreement with both the occupation fraction expected from the Fermi surface of a monovalent metal and the experimental fractions determined from Compton~\cite{billington2015magnetic}, ARPES~\cite{Sobota.2013}, and quantum oscillations~\cite{Ok.2013}. 

The DFT projected occupancy has some similarities to the experiment around $\overline{K}$. This feature in the DFT relates to the Cr DFT band crossing the Fermi level near $K$ (and the corresponding points along the $k_z$-axis) resulting in an electron Fermi surface pocket around $K$ (see Fig.~\ref{dmftband}~(b) and (c)). However, the agreement at $\overline{K}$ significantly worsens in the DFT+DMFT (at $J = 0.7$ eV) as there is no contribution from the Cr band as it is now below the Fermi level and hence fully occupied (insulating). Interestingly, however, there is a small contribution at $\overline{K}$ in the DFT+DMFT ($J = 0.7$ eV) projected occupations which arises from the spectral weight of the Pd quasiparticle conduction band spilling across the Fermi level (such as that seen around $K$ in Fig.~\ref{dmftband}~(d)) which then becomes occupied, similar to that seen in Refs.~\cite{PhysRevLett.124.046402,Chen.2022} as discussed previously. This additional spectral weight is small relative to the background (i.e., relative to the $\Gamma$ point) which would likely mean that this feature might be difficult for the ARPES to distinguish within the experimental and statistical error. It should be noted that the projected occupation from Compton presented here relates to the energy integral of the occupied part of the spectral function which is then integrated along the $k_z$-axis. Consequently, the accumulation of this feature around $K$ seen in the spectral function in Fig.~\ref{dmftband}~(d) becomes more prominent in the projected occupation at $\overline{K}$. This is seen in the DFT+DMFT ($J = 0.7$ eV) projected occupation feature around $\overline{K}$ in Fig.~\ref{trans_occ}. We note that the DFT+DMFT spectral plot along the same in-plane path as Fig.~\ref{dmftband}~(b) but with a shift of 0.5 reciprocal lattice units along the $k_z$-axis shows a similar dispersion to $k_z = 0$ plane which is expected for this quasi-2D system. Therefore, this DFT+DMFT feature at $\overline{K}$ will have contributions from all the spectral weight spillage along the $k_z$-axis centred at $K$ owing to the projected nature of the Compton occupation data.  

Also shown in Fig.~\ref{trans_occ} are several DFT+DMFT calculations of the 2D projected occupancy plotted for different $J$ but with  $U$ fixed to 3.0 eV. These show the evolution of the 2D projected occupancy (and by inference, the electronic structure) as a function of the size of the Hund exchange interaction $J$ as the CrO$_2$ layer transitions from the metallic (low $J$) to Mott insulating state (high $J$), where the Mott insulating state occurs for $J > 0.25$ eV. The result of increasing $J$ causes the smoothed double-step feature prominent in the DFT projected occupancy along $\overline{\Gamma}$ to $\overline{M}$ to transform into a single step due to the spectral weight from the previously conducting Cr quasiparticle bands shifting below the Fermi level and becoming fully occupied. On the other hand, increasing $J$ suppresses the 2D occupancy contribution around $\overline{K}$ as the Cr quasiparticle bands transition into being Mott insulating. There are no optimal DFT+DMFT $U$ and $J$ parameters which are able to simultaneously capture the 2D projected occupancy features from $\overline{\Gamma}$ to $\overline{K}$ and the single step along $\overline{\Gamma}$ to $\overline{M}$. The hexagonal Fermi surface sheet is a robust feature in all the Fermi surface measurements and is clearly captured by the DFT+DMFT predictions with Mott insulating CrO$_2$ layers.

\begin{figure}[t!]
 \centerline{\includegraphics[width=\linewidth]{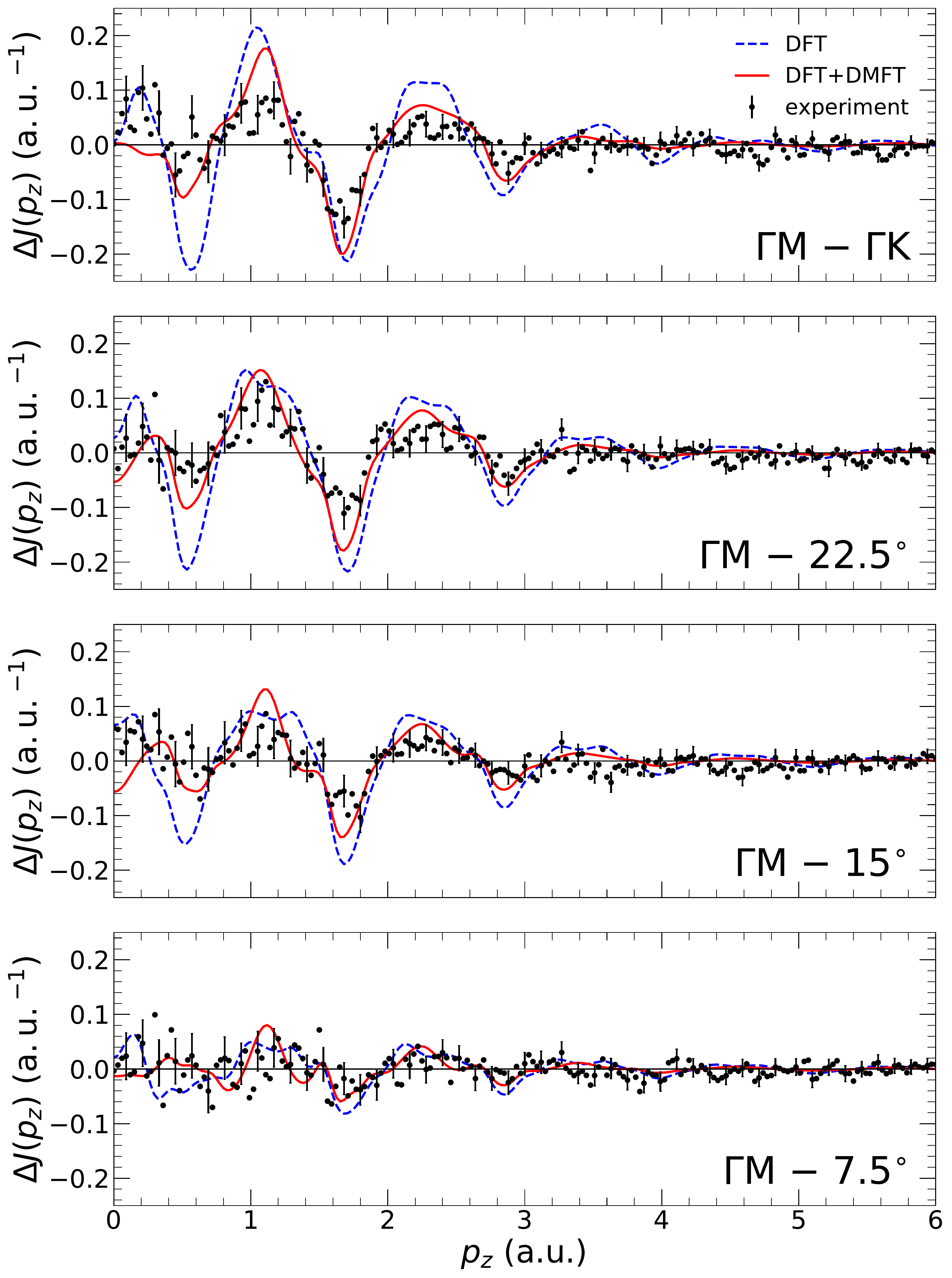}}
 \caption{The directional differences $\Delta J(p_z)$ (i.e., the difference between two Compton profiles measured along different crystallographic directions) as specified at the bottom right of each panel where the angle refers to the rotation away from the $\Gamma M$ direction towards $\Gamma K$. These differences are of the DFT, DFT+DMFT, and the experiment. The theoretical Compton profiles were convoluted with a one dimensional (1D) Gaussian of full-width-at-half-maximum $=0.106$ a.u. to represent the experimental momentum resolution. The experimental data are from Ref.~\cite{billington2015magnetic}.} 
\label{Jp}
\end{figure}

\begin{figure}[t!]
\centerline{\includegraphics[width=\linewidth]{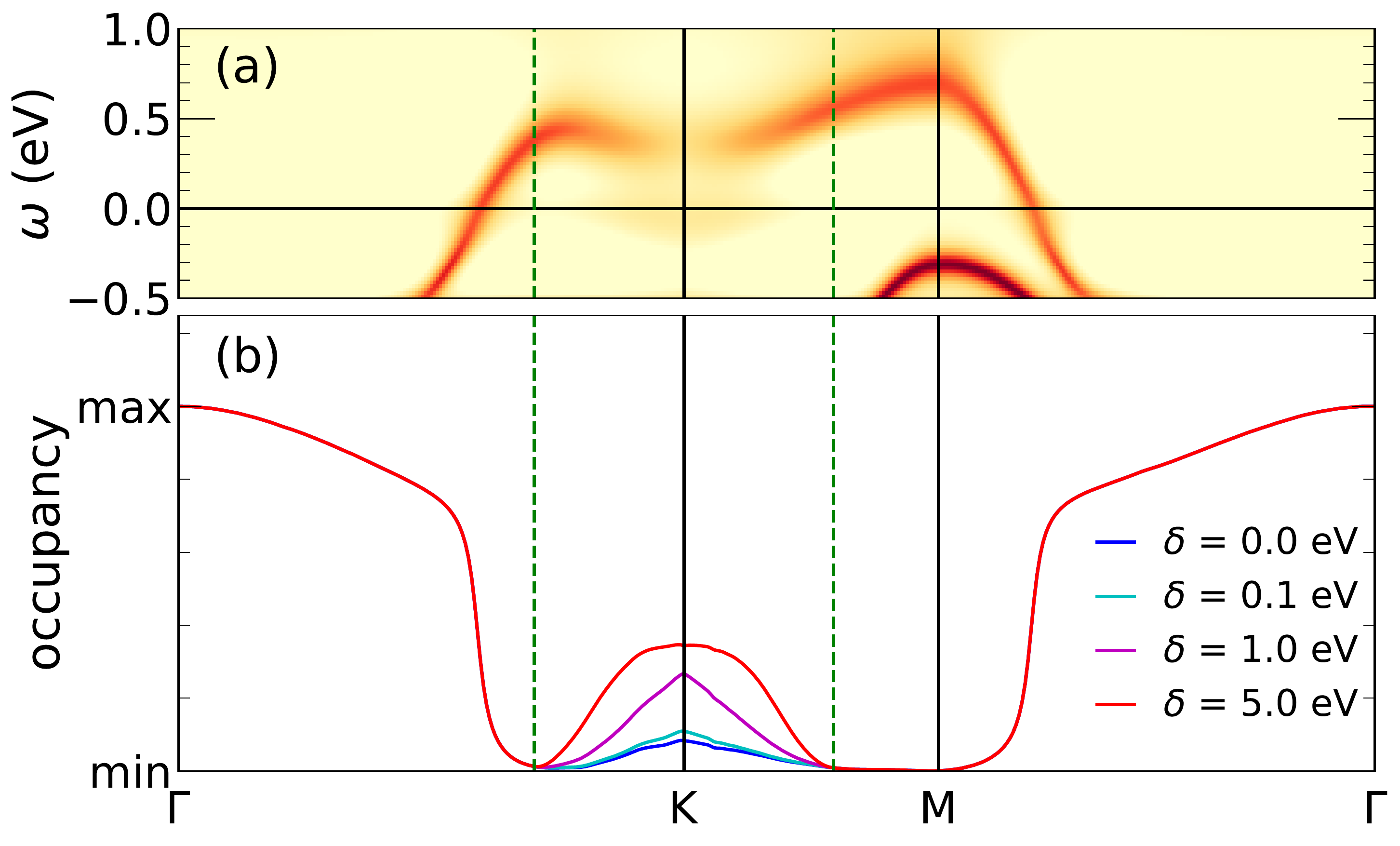}}
\caption{(a) The logarithm of DFT+DMFT spectral function with an additional artificial broadening term (in energy) along the same path and colour scale as in Fig.~\ref{dmftband}~(b). This broadening is only applied to the Pd quasiparticle conduction band centred around $K$ up to the dashed boundaries. This broadening term varies quadratically from zero at the dashed boundaries to a maximum of $\delta$ (here it is equal to 1 eV) at $K$. There is no physical significance to the relation between the additional broadening and its $\mathbf{k}$-dependence, it just ensures a continuous change in the broadening. (b) The occupancy along this path obtained from integrating the artificially broadened spectral function up to the Fermi level for different maximum $\delta$ values given in the legend. Both panels help to show how the spectral function (measured by ARPES) and occupancy (measured by Compton scattering) are related to each other, along with the different features of the electronic structure ARPES and Compton scattering would probe.} 
\label{shift_occ}
\end{figure}

To get a better perspective of the agreement between the different calculations with the experimental data, Fig.~\ref{2docc} shows the 2D projected occupancy of the different calculations and experiment. The DFT results give good agreement in certain regions, but is overall worse than the DFT+DMFT as expected. The size of DFT+DMFT hexagonal occupancy weight around $\overline{\Gamma}$ is in excellent agreement with the experimental 2D projected occupancy, as previously established. However, it is clear that the DFT+DMFT is unable to predict the significant additional occupation feature surrounding the hexagonal region which gives rise to elongated black ellipsoidal region centred around $\overline{M}$, with the major axis of this ellipsoid along the $\overline{M}$---$\overline{K}$ path. Next, we present the comparison of the directional differences along the different measured (crystallographic) directions in Fig.~\ref{Jp} for the DFT, DFT+DMFT and the experimental data. It is clear that the DFT+DMFT results are superior in agreement with the experiment compared with the DFT. 

Thus far, the origin of the features measured in the experimental techniques has been discussed from a theoretical perspective. However, the discrepancy between experimentally measured features by ARPES and Compton still needs to be addressed. Both of these experiments were performed at different temperatures with the Compton being at room temperature, whereas the ARPES was measured at 100 K. There have been no reported signatures which could be related to a temperature-dependent Lifshitz transition in the transport measurements~\cite{Takatsu.2010.2} which could have explained this extra feature in the Compton data at $\overline{K}$ being related to the Fermi surface. However, it should be noted that the spectral weight of the Pd quasiparticle conduction band would be more broadly distributed in energy in the room temperature Compton data than the 100 K ARPES data meaning more spectral weight from the tail of that quasiparticle band would likely be occupied. It would be strange if the ARPES spectra would miss a Fermi surface feature at $K$ due to cross-section effects as it is very unlikely for ARPES not measure the same band in different regions of the Brillouin zone. It is also unlikely that the ARPES matrix elements effects are suppressing a Fermi surface feature originating from the Pd quasiparticle band, although ARPES matrix elements effects do cause some changes in the measured intensity~\cite{Sunko.2020}. The reduced dimensionality at the surface may enhance the electron correlation effects within the Mott insulating CrO$_2$ layers at the surface, similar to that seen in SrVO$_3$~\cite{maiti2001,liebsch2003,laverock2015b,ishida2006,yoshimatsu2010,zhong2015}. On the other hand, there is unlikely any notable contribution from surface states in the ARPES as these would give additional features \cite{Sobota.2013}, not remove some.

Returning to the experimental feature at $\overline{K}$, one possible explanation is that this may actually arise from the DFT+DMFT Pd conduction quasiparticle band at $K$ (and the positions displaced along $k_z$) being broader and/or closer to the Fermi level than predicted in the DFT+DMFT with the feature arising from the spectral weight spillage. A computationally inexpensive way to gain some insight into the contribution that this part of the quasiparticle band would make to the occupancy is to add an artificial (and arbitrary) broadening term (in energy) to this DFT+DMFT Pd quasiparticle conduction band around the $K$ point as shown in Fig.~\ref{shift_occ} (a). It is clear how dispersive this makes the quasiparticle band around $K$ resulting in additional spectral weight spillage crossing the Fermi level which gives rise to a more prominent occupancy feature at $K$ in Fig.~\ref{shift_occ} (b). The occupancy in Fig.~\ref{shift_occ} (b) is calculated by integrating the real-frequency-dependent broadened spectral function up to the Fermi level. This feature grows as a function of increasing $\delta$, which is the maximum of the additional broadening as explained in Fig.~\ref{shift_occ} (b), until exceeding $\delta=5$ eV. The occupation for the unbroadened ($\delta=0.0$ eV) spectral function is very similar to the DFT+DMFT ($J=0.7$ eV) 2D projected occupancy in Fig.~\ref{trans_occ}, with the additional smearing in that occupancy coming from the convolution with the experimental momentum resolution function. This similarity is to be expected as this is a quasi-2D electronic structure. We note that the additional occupation from this broadening violates charge conservation and as such, the Fermi level would need to move to compensate for this. 

This broadened spectral function serves to illustrate how the feature at $\overline{K}$ in the experimental Compton data may arise from this quasiparticle band. However, even with the unphysical arbitrary broadening, it is still not enough to fully agree with the experimental Compton data. This would suggest that the shape of this quasiparticle band may need to change with the dip around $K$ likely being closer to the Fermi level, but its band centre must remain above the Fermi level to agree with the established single hexagonal Fermi surface. We emphasise Fig.~\ref{shift_occ} illustrates how ARPES and Compton probe the electronic structure differently, in this case around $K$. For $\delta=1$ eV, Compton scattering would probe a distinct occupation feature around $K$, but the spectral function at the Fermi level around $K$ is relatively small in magnitude which may make it difficult to distinguish in ARPES. We note that Lechermann~\cite{Lech.2018} showed that the introduction of relatively large (electron) doping results in a downward shift in energy of the Pd quasiparticle band around $K$. This will likely give a more prominent feature in the occupancy feature around $K$, for the reasons previously discussed. However, the PdCrO$_2$ single-crystal sample measured by Billington $\textit{et al.}$ were grown by H. Takatsu as described in Ref.~\cite{TAKATSU20103461} and were of similar high purity and quality to those measured by ARPES~\cite{Sobota.2013,Noh.2014,Sunko.2020} and quantum oscillations~\cite{Ok.2013,PhysRevB.92.014425}. Therefore, it is highly unlikely that the measured feature at $\overline{K}$ in the projected occupation comes solely from naturally occurring doping effects, but their contributions cannot be fully ruled out.

The Cr $3d$ DMFT self-energy significantly influences the Pd quasiparticle conduction band around $K$ due to coupling between the layers of the localised Cr and itinerant Pd electrons, as discussed in detail by Lechermann~\cite{Lech.2018}. This type of coupling is similar to the Kondo effect, but here the localised spins in PdCrO$_2$ originate from a Mott mechanism which suppresses the electron hopping between sites. The inclusion of the DMFT self-energy brings the Pd quasiparticle conduction band closer to the Fermi level around $K$ and redistributes a significant amount of the Cr $3d$ contribution to the spectral function away from this quasiparticle band peak and into the Hubbard-like bands. The disagreement in the occupancy may stem from inadequacies in the description of the hybridisation between the Cr $3d$ and Pd $4d$ states (which relates to the inter-layer electron coupling) at the DFT level. This can be very sensitive to the exchange-correlation functional used at the DFT level, as seen for group V and VI elements~\cite{PhysRevB.103.235144}. Considering that the Pd states are primarily treated on the DFT level, higher order electron correlation contributions may influence the Pd conduction quasiparticle band dispersion and impact the inter-layer electron coupling. The correct description of this inter-layer coupling may cause the shape and broadening of the Pd quasiparticle band to change to give the (2D projected) occupancy feature at $\overline{K}$ revealed by Compton scattering while also being potentially difficult to distinguish in ARPES. We note that the reduced dimensionality at the surface could influence the inter-layer electron coupling (and other electron correlation effects), which may alter the Pd quasiparticle conduction band shape and dispersion which ARPES (potentially) would measure in comparison to what the Compton scattering bulk-probe measures. The results of DFT+DMFT calculations performed with an additional impurity site for the Pd $4d$ orbitals (with the Cr and Pd DMFT impurities are treated independently) do not significantly alter the presented DFT+DMFT results which suggests that the local Pd electron correlations are insignificant when it comes to explaining the origin of the missing feature around $\overline{K}$ in the projected occupations.

There is increasing amount of experimental evidence showing significant inter-layer electron coupling. Transport measurements in Ref.~\cite{Takatsu.2010.2} show that the frustrated Cr spins affect the out-of-plane and in-plane motion of the conduction electrons in the Pd layer. The interpretation of the magnetothermopower measurements in Ref.~\cite{PhysRevLett.116.087202} also point to there being significant coupling between the itinerant Pd electrons with the short-range electron spin-correlations of the Cr electron spins well above $T_{\rm N}$. The short-range electron spin correlations which persisted above $T_{\rm N}$ were also measured by single-crystal neutron scattering in Ref.~\cite{billington2015magnetic}. Further transport measurements have shown the effect of the short-range order on the Hall and Nernst effects~\cite{PhysRevB.92.245115}. Raman and electron spin resonance (ESR) measurements~\cite{Glamazda.2014} have also shown evidence for inter-layer hoppings along the $c$-axis and a reconstruction of electronic bands on approaching $T_{\rm N}$. Recent ARPES~\cite{Sunko.2020} measurements in the antiferromagnetic phase showed that the measured spectra can be explained by an intertwined excitation consisting of a convolution of the charge spectrum of the metallic Pd layer and the spin susceptibility of the Mott insulating CrO$_2$ layer. This excitation arises from an inter-layer Kondo-like coupling. The authors of Ref.~\cite{Sunko.2020} draw parallels with the results of the doping calculations of the Mott layer calculated in Ref.~\cite{Lech.2018} which, as already discussed, significantly affects the shape and dispersion of the Pd quasiparticle band at $K$. They emphasise that the results of their measurements and the doped DFT+DMFT calculations reflect the fact that in a coupled Mott-itinerant system, the itinerant layer will support charge excitations~\cite{Sunko.2020}. As the short-range electron spin correlations persist beyond $T_{\rm N}$, our interpretation of the Compton results with respect to the Pd quasiparticle band ties in with the experimental evidence of the inter-layer electron coupling, and may be linked to the intertwined excitation. Therefore, electron correlation effects which contribute to the inter-layer electron coupling, such as those in the models used in Refs.~\cite{Sunko.2020,McRoberts}, beyond those included in our DFT+DMFT calculations, seem to be significant. To confirm that the Pd conduction quasiparticle band is indeed broader and closer to the Fermi level than that predicted, the experimental $\textbf{k}$-resolved dispersion of that band could, for example, be measured by pump-probe ARPES or $\textbf{k}$-resolved inverse photoemission spectroscopy (KRIPES) experiments which can probe the unoccupied part of the band structure.

\section{Conclusion}
We have shown that the paramagnetic DFT+DMFT theoretical description of the electronic structure of PdCrO$_2$ is superior to DFT as it gives excellent agreement with the features relating to the hexagonal Fermi surface sheet measurement by all the Fermi surface experimental data, all of which agrees with the picture of the Mott insulating CrO$_2$ layers~\cite{Mackenzie_2017}. However, there are still discrepancies between the paramagnetic DFT+DMFT results and the Compton data measured within the paramagnetic phase. We found that there is no combination of $U$ and $J$ around the Mott insulator transition (in the CrO$_2$ layers) in DFT+DMFT which agrees with the presence of both the hexagonal Fermi surface and the feature around $\overline{K}$ as measured by the Compton. By adding an unphysical broadening term (in energy) to the DFT+DMFT the Pd quasiparticle conduction band around $K$, more spectral weight spills across the Fermi level which gives rise to a more prominent feature in the occupancy. However, this is still not enough to agree with the measured projected occupancy feature in the Compton data, so a change in both the broadening and shape of this quasiparticle band is needed while keeping its band centre above the Fermi level to avoid any changes to the established Fermi surface topology. Overall, our DFT+DMFT results help to clarify the origin of features in the Compton data.

From the available experimental and theoretical evidence thus far, the feature in the projected electron occupancy measured at $\overline{K}$ by Compton scattering is likely from the spectral weight of the Pd conduction quasiparticle band spilling across the Fermi level and becoming occupied. The ARPES may not measure this proposed spectral weight spillage if the Pd quasiparticle band is very dispersive around $K$ (and the positions displaced along $k_z$) and if the surface influences the electron correlation effects, such as the inter-layer electron coupling, which may then alter the quasiparticle band shape and dispersion. As the DFT+DMFT model used does not predict the measured projected occupation feature at $\overline{K}$, theories beyond our DFT+DMFT are required to establish the exact origin of this feature, which likely relates to the inter-layer electron coupling between the Pd and CrO$_2$ layers which gives rise to new Kondo-like physics such as the previously observed intertwined excitation~\cite{Sunko.2020}. The discrepancy with the Compton data gives motivation to experimentally measure the dispersion of the unoccupied part of the Pd quasiparticle conduction band to determine if it is indeed closer to the Fermi level and much more smeared in energy than predicted by our DFT+DMFT calculations. Evidently, Compton scattering is a powerful probe of many-body electron correlation effects.

\section{Acknowledgements}
A.D.N.J. acknowledges the Doctoral Prize Fellowship funding and support from the Engineering and Physical Sciences Research Council (EPSRC). We are grateful for the useful discussions with J. Laverock, M. Favaro-Bedford, Wenhan Chen, and C. Mackellar. Calculations were performed using the computational facilities of the Advanced Computing Research Centre, University of Bristol (\href{http://bris.ac.uk/acrc/}{http://bris.ac.uk/acrc/}). The VESTA package (\href{https://jp-minerals.org/vesta/en/}{https://jp-minerals.org/vesta/en/}) has been used in the preparation of some figures.

\end{document}